\documentclass[fleqn,10pt,twocolumn]{wlscirep}

\usepackage[utf8]{inputenc}
\usepackage[T1]{fontenc}

\newcommand{\STO}{SrTiO$_3$}
\newcommand{\LAO}{LaAlO$_3$}
\newcommand{\ETO}{EuTiO$_3$}
\newcommand{\CSTO}{Sr$_{0.99}$Ca$_{0.01}$TiO$_3$}

\newcommand{\etal}{\textit{et al.}}
\makeatother
\usepackage{svg}
\usepackage{multicol}
\usepackage{graphicx}
\usepackage{blindtext}
\usepackage{caption}
\usepackage{dblfloatfix} 
\usepackage{xcolor}
\usepackage{float}
\usepackage{gensymb}

\begin{document}
\title{Tunable Magnetic Scattering and Ferroelectric Switching at the \LAO/\ETO/\CSTO~Interface}

\author[1]{Gal Tuvia}
\author[2]{Sapir Weitz Sobelman}
\author[1]{Shay Sandik}
\author[2,*]{Beena Kalisky}
\author[1,*]{Yoram Dagan}
\affil[1]{School of Physics, Tel Aviv University, Tel Aviv 6997801, Israel}
\affil[2]{Department of Physics and Institute of Nanotechnology and Advanced Materials, Bar-Ilan University, Ramat-Gan 5290002, Israel}
\affil[*]{Beena@biu.ac.il , Yodagan@tauex.tau.ac.il}

\begin{abstract}
Ferroelectric and ferromagnetic orders rarely coexist, and magnetoelectric coupling is even more scarce. A possible avenue for combining these orders is by interface design, where orders formed at the constituent materials can overlap and interact. Using a combination of magneto-transport and scanning SQUID measurements, we explore the interactions between ferroelectricity, magnetism, and the 2D electron system (2DES) formed at the novel \LAO/\ETO/\CSTO~heterostructure. We find that the electrons at the interface experience magnetic scattering appearing along with a diverging Curie-Weiss-type behaviour in the \ETO~layer. The 2DES is also affected by the switchable ferroelectric polarization at the \CSTO~bulk. While the 2DES interacts with both magnetism and ferroelectricity, we show that the presence of the conducting electrons has no effect on magnetization in the \ETO~layer. Our results provide a first step towards realizing a new multiferroic system where magnetism and ferroelectricity can interact via an intermediate conducting layer.\end{abstract}
\maketitle
\section*{Introduction}

\begin{figure*}[!ht]
\centering
  \includegraphics[scale=0.1] {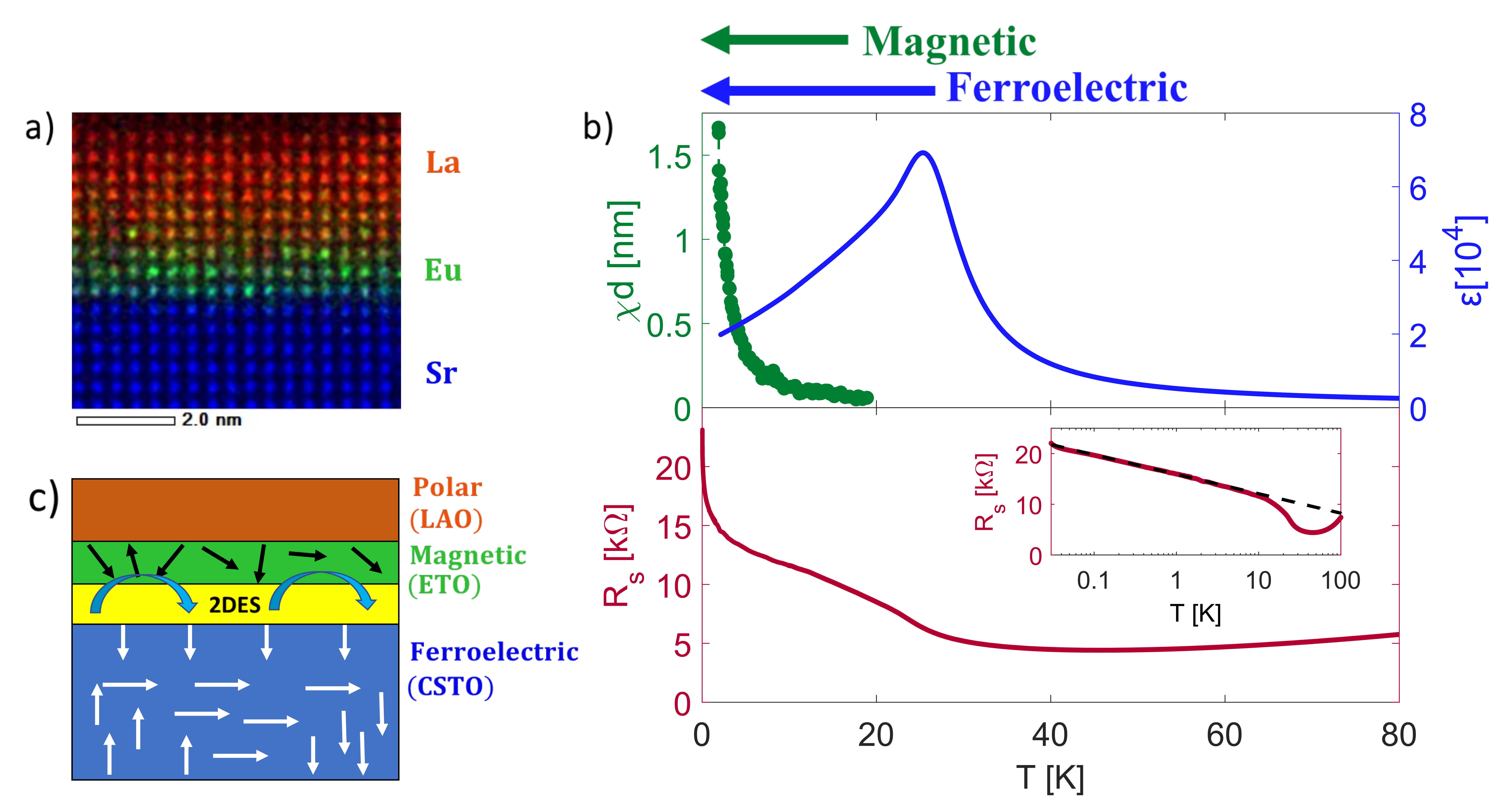}
  \caption{\textbf{Magnetism and ferroelectricity interacting with the 2DES} (a) STEM measurement along with EELS elemental identification confirming the high quality growth of the interface. (b) Top panel:  Green (left scale): Magnetic susceptibility versus temperature, showing a Curie-Weiss behaviour originating from the ETO layer. Blue (right scale): Dielectric constant of the CSTO substrate. A ferroelectric transition is observed at $\sim$30K. Bottom panel: Interface sheet resistance versus temperature. Inset: Sheet resistance versus temperature in logarithmic scale, highlighting the $ln(\frac{1}{T})$ behaviour at low temperatures. (d) Sketch of our understanding of the system: The 2D polar metal is controllable by switching the bulk ferroelectricity and experiences scattering from spin fluctuations in the ETO layer.} 
  \label{fig:Mainplot}
\end{figure*}

Materials exhibiting ferroelectricity and ferromagnetism are long-sought because of their potential use in multifaceted memory devices. However, long-range orders such as ferroelectricity, ferromagnetism, and superconductivity rarely coexist. For example, ferroelectricity often requires empty \textit{d}-orbitals, while conventional ferromagnets require partially filled \textit{d}-orbitals\cite{hill2000there}. Superconductivity and ferroelectricity also seem exclusive due to the inversion symmetry breaking needed for the latter\cite{matthias1967ferroelectricity}. A possible avenue for combining seemingly incompatible orders is by interface design, where orders formed at the constituent materials can overlap. This approach has been successful in combining ferroelectricity and ferromagnetism by designing composite heterostructures\cite{mundy2016atomically}. Furthermore, a two-dimensional superconductor was recently realized at the surface of a ferroelectric Ca-substituted \STO~ crystal\cite{tuvia2020ferroelectric}. The incorporation of ferroelectricity to conducting interfaces raises the possibility of controlling other properties, such as magnetism by ferroelectric switching.

In its bulk form, \ETO~(ETO) is a G-type antiferromagnet with a  N\'eel temperature of 5.5K\cite{mcguire1966magnetic,chien1974magnetic,scagnoli2012eutio}. However, when strained\cite{lee2010strong} or doped\cite{takahashi2009control,shimamoto2013full,kususe2014magnetic,katsufuji1999transport,yamamoto2015antiferro,ahadi2017evidence,kugimiya2007preparation} it becomes ferromagnetic with Curie temperatures ranging from 4K-12K. The emergence of a ferromagnetic phase by straining is explained by a change of magnetic exchange parameters\cite{ryan2013reversible}. Straining also drives ETO into a ferroelectric phase\cite{lee2010strong}, making it a candidate system for multiferroic research. Doping-induced magnetism on the other hand, is believed to result from Ruderman–Kittel–Kasuya–Yosida (RKKY) interactions\cite{yamamoto2015antiferro,kugimiya2007preparation,ahadi2017evidence,takahashi2018anomalous}.

ETO is isomorphic to \STO~(STO). It has been shown that one-or two-unit cells of ETO can be grown between STO and \LAO~(LAO) without destroying the conducting interface\cite{stornaiuolo2016tunable}, which is well known to form at the LAO/STO interface\cite{ohtomo2004high}. The resulting LAO/ETO/STO heterostructure displays gate tunable superconductivity and non-linear Hall resistivity below 10K suggested to originate from a tunable ferromagnetic phase\cite{stornaiuolo2016tunable}.

Here, we report magneto-transport of the \LAO/\ETO/\CSTO~(LAO/ETO/CSTO) heterostructure combined with local mapping of electrical currents and magnetism using scanning superconducting quantum interference device (SQUID) microscopy. We find a diverging Curie-Weiss-type susceptibility at low temperatures appearing along with with a $ln(\frac{1}{T})$ term in the sheet resistance. We interpret the magneto-electric-transport properties in the framework of a 2D polar metal affected by scattering off magnetic fluctuations. Ferroelectricity also interacts with the 2DES as demonstrated by the hysteresis observed in sheet resistance when switching ferroelectric polarization in the bulk. Our results demonstrate that a 2D electron system can be designed to interact with ferroelectricity and magnetism. While the 2DES is affected by both the bulk ferroelectiricty and the magnetic ETO layer, we find that the presence of the conduction electrons has no effect on the ETO magnetism.

\section*{Results}
\textbf{Heterostructure design:} Our goal in this work is to realize a conducting interface exhibiting both ferroelectric and magnetic properties. To design such a system we chose three components: 
\begin{enumerate}
    \item Ferroelectric CSTO, which will serve as the substrate for the heterostructure
    \item A thin two-unit cell layer of magnetic ETO
    \item Eight-unit cells of LAO, which induce conductivity at the interface
\end{enumerate}

The resulting heterostructure can be seen in figure \ref{fig:Mainplot} (a), where we show Scanning-Tunneling-Electron-Microscopy (STEM) imaging  along with Electron-Energy-Loss-Spectroscopy (EELS) measurements, verifying the expected structure and composition of our heterostructure (further details are shown in supplementary figure S1).

The interface is metalic at high temperatures (Figure \ref{fig:Mainplot} (b), bottom panel, red). Upon further cooling of the heterostructure, two distinct features can be observed - an upturn to the resistance below the ferroelectric transition, and a $ln(\frac{1}{T})$ term becoming dominant at low temperatures (highlighted in inset of bottom panel). As we elaborate below, we relate both of these terms to interactions of the 2D electron system (2DES) with the ferroelectric bulk and the magnetic ETO layer. 

\textbf{Ferroelectricity:} We begin by characterizing ferroelectricity in the bulk CSTO crystal. The dielectric constant is extracted by capacitance measurements and is presented in Figure \ref{fig:Mainplot} (b) (top panel, blue). A ferroelectric transition is observed at $\sim$30K, as expected for one percent Ca substitution\cite{bednorz1984sr}. 

\begin{figure}[!ht]
\captionsetup{type=figure}
    \includegraphics[width=\columnwidth]{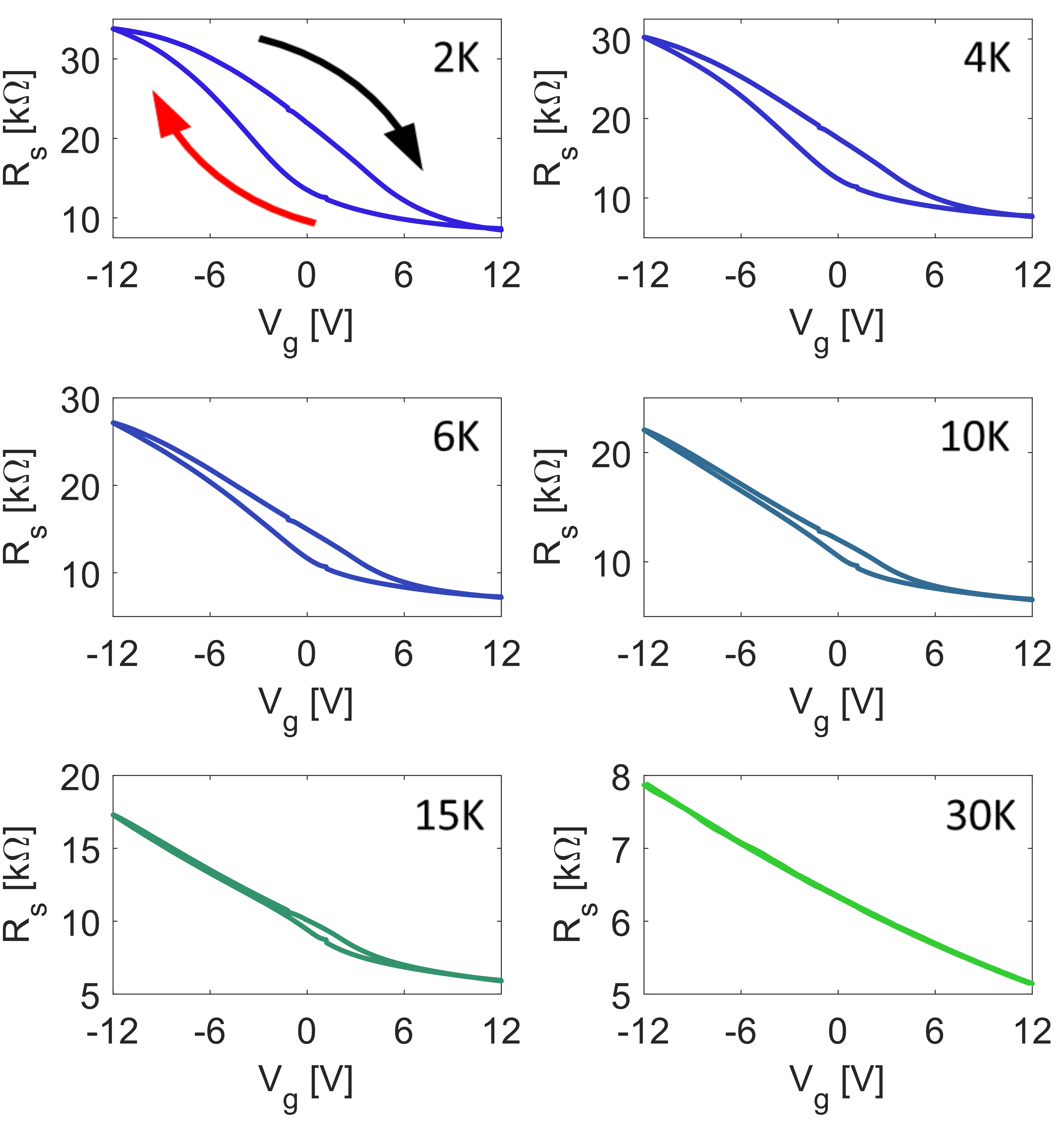}
    \caption{\textbf{Temperature Evolution of Resistance Versus Gate Hysteresis} Sheet resistance response to applied back gate voltage shows hysteretic behaviour below the ferroelectric transition. This hysteresis becomes larger as temperature is lowered deeper into the ferroelectric phase, demonstrating control of interface resistance by ferroelectric switching.}
    \label{fig:RVG}
\end{figure}

The ferroelectric transition has a dramatic effect on the interface resistance. As can be seen in \ref{fig:Mainplot} (b) (bottom panel), the interface resistance increases as temperature is lowered into the ferroelectric phase. We have previously interpreted similar upturn in resistance as a result of the interaction between the ferroelectric bulk and the conducting polar interface\cite{tuvia2020ferroelectric}. Furthermore, as temperature is lowered deeper into the ferroelectric phase, resistance versus back-gate voltage hysteresis loops are observed (Figure \ref{fig:RVG}), demonstrating ferroelectric control of the 2DES. The area of these loops increases as temperature is lowered, as previously reported for 2D polar metals formed at CSTO-based interfaces\cite{tuvia2020ferroelectric,brehin2020switchable}.

\begin{figure}[!ht]
\captionsetup{type=figure}
\includegraphics[scale=0.45]{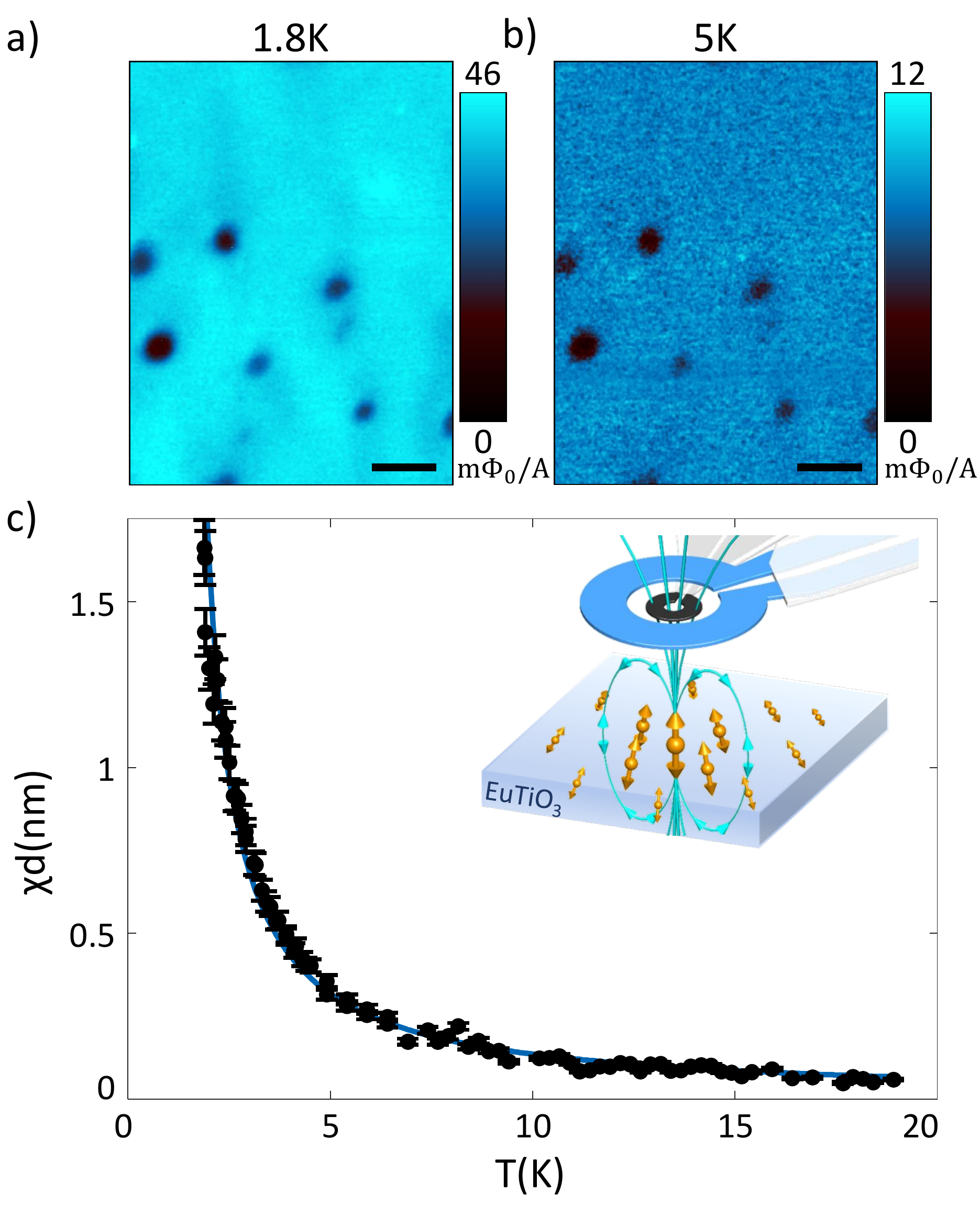}
\caption{\textbf{The paramagnetic-type signal decays with the temperature.} (a-b) Maps of the paramagnetic-type response at 1.8K (a) and 5K (b). Scale bars 20µm; (c) Paramagnetic susceptibility as a function of the temperature. Inset: Illustration of the SQUID susceptometry measurement.}
    \label{fig:SQUID}
\end{figure}

\textbf{Magnetism:} In figure \ref{fig:SQUID} (a)-(b) we present spatial susceptibility measurements of our interface at 1.8K and 5K respectively. The signal is mostly isotropic and only changes at imperfections on the interface (black regions in figure \ref{fig:SQUID} (a)-(b)), where the magnetic signal, as well as the conductivity, are substantially weaker (see supplementary figure S2 for more details). 

To quantify the evolution of the magnetic signal with temperature, we perform a Curie-Weiss fit to the susceptibility measurements (solid line in figure \ref{fig:SQUID} (c)):

\begin{center}
$\frac{1}{\chi d}=\frac{3K_B}{\mu_0(g\mu_B)^2J(J+1)}\frac{1}{n_s}(T-T_C)$
\end{center}

Where $\chi$ is unit-less susceptibility, $d$ is the thickness of the magnetic layer, $J$ is the nuclear spin number, $n_s$ is the spin density and $T_C$ is the Curie temperature.

We can now extract the spin density $n_s$ from the fit to find the origin of the magnetic response. If we assume $J=\frac{7}{2}$ (i.e. that the magnetic signal originates from the Eu atoms) we receive a spin density of $n_s=8.2\cdot10^{14} \frac{1}{cm^2}$, which is the expected order of the number of Eu atoms. However, if we assume $J=\frac{1}{2}$ (i.e. that the magnetic signal originates from the 2DES), we receive a spin density of $n_s=1.7\cdot10^{16} \frac{1}{cm^2}$, which can be ruled out since it is two orders of magnitude larger than the overestimated carrier density of half an electron per unit cell\cite{ohtomo2004high}.

We note that the spin densities calculated here are lower limits as the susceptibility was measured with respect to the imperfections on the sample, which have a much weaker yet non-zero signal. By subtracting the background signal away from the sample we can set a more realistic estimation for the signal of $1.4\cdot10^{15} \frac{1}{cm^2}$ for $J=\frac{7}{2}$. However, this estimation is hindered by noise (experimental limitations, see supplementary figure S3 for more information). We therefore conclude that for $J=\frac{7}{2}$, the spin density is larger than $8.2\cdot10^{14}$, likely around $1.4\cdot10^{15} \frac{1}{cm^2}$. This value is in excellent agreement with the expected density of Eu atoms within two-unit cells of ETO ($1.3\cdot10^{15}\frac{1}{cm^2}$), further confirming the magnetic signal originates from the ETO layer.

The Curie-Weiss behaviour persists down to the 1.7K (our SQUID experimental base temperature), with no clear signs of a magnetic phase transition. Interestingly, the $ln(\frac{1}{T})$ term in the interface resistance is observed at the same temperature range. We interpret this behaviour as a signature of magnetic fluctuations in the ETO layer scattering conduction electrons in the 2DES. This view is consistent with the magnetoresistance (MR) measurements showed below. 

\begin{figure*}[!ht]
\centering
\captionsetup{type=figure}
  \includegraphics[scale=0.15] {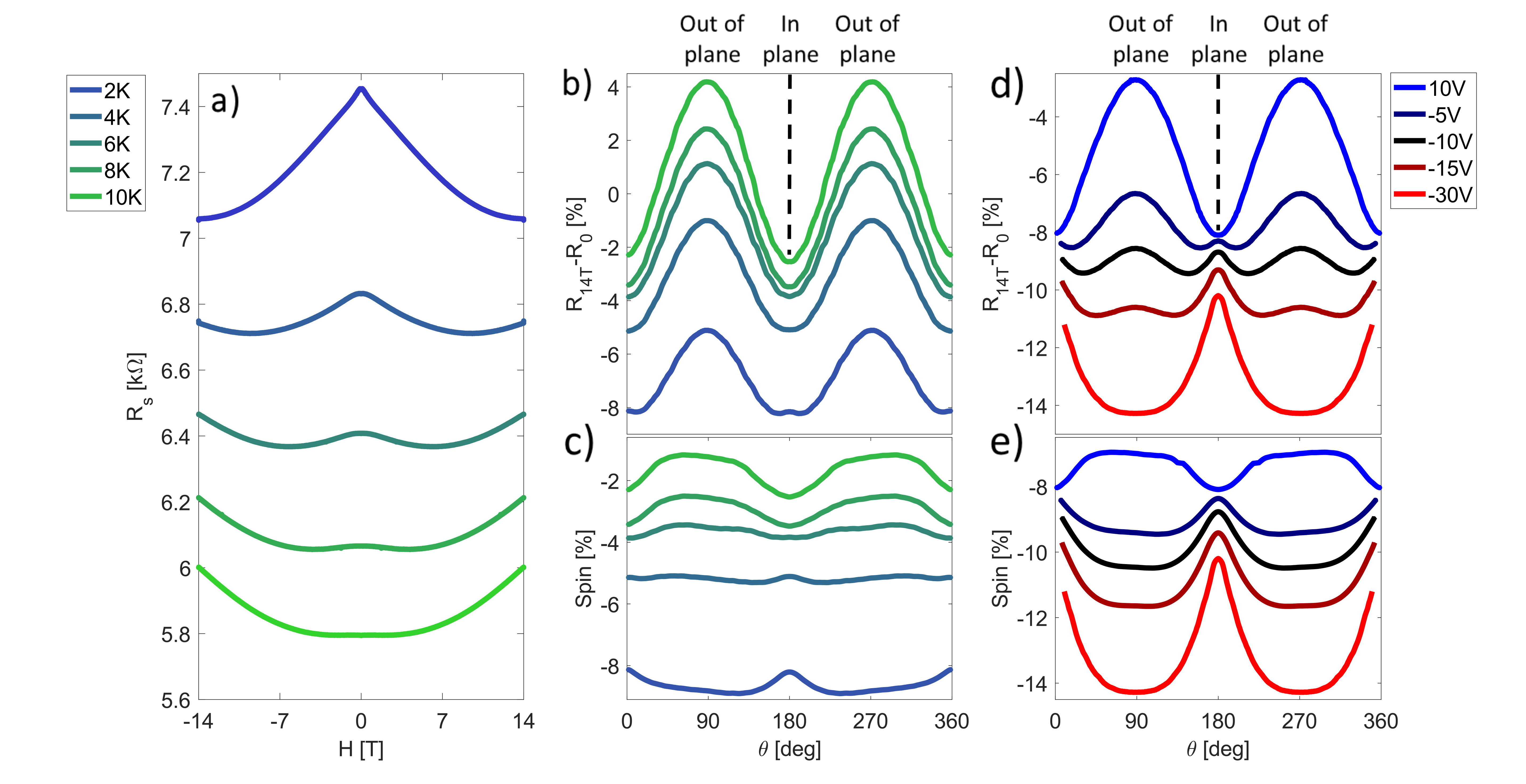}
  \caption{\textbf{Low Temperature Magnetoresistance (MR) Measurements} (a) Symmetrized resistance versus out-of-plane magnetic field for different temperatures. (b) MR magnitude at 14T for different temperatures and angles between the interface and the field ($\theta$). (c) MR magnitude after subtraction of extrapolated orbital effect. This remaining effect illustrates the MR spin component. (d) MR magnitude at 2K for different back-gate voltages and angles $\theta$. (e) Gate-dependant spin contribution achieved similarly to (c). This term becomes more dominant at negative gates (electron depletion).} 
  \label{fig:MR}
\end{figure*}

In figure \ref{fig:MR} (a), we present resistance versus magnetic field measurements with field perpendicular to the interface at different temperatures. The sign of the MR depends on both field magnitude and temperature. We interpret this behaviour as a result two competing effects: a negative spin MR, resulting from interactions with the magnetic ETO layer, and a positive, orbital quadratic term. At low temperatures, the spin component is stronger while at sufficiently high magnetic fields, the quadratic component dominates.

To quantify the spin and orbital contributions, we rotate the sample at a constant magnetic field of 14T. The resulting MR($\theta$) measurements are presented in figure \ref{fig:MR} (b), where $\theta$ is the angle between the field and the interface. For $70\degree\leq\theta\leq110\degree$, we assume that the magnetic anisotropy is negligible, hence we fit the data in this regime to $\alpha+\beta\cdot Sin^2(\theta)$, where $\alpha$ and $\beta$ are constants representing the magnitude of the spin and orbital effects respectively. The resulting orbital component is presented in supplementary figure S8. We then subtract this positive component from the original signal to obtain the spin contribution (figure \ref{fig:MR} (c)). We find that the spin component is negative and becomes stronger as the temperature is lowered, as expected for magnetic fluctuations increasing at low temperatures. Furthermore, we observe an anisotropy of the spin component as $\theta$ approaches zero (in-plane magnetic field), reflecting the 2D nature of the ETO layer.

The magnitude of both spin and orbital components can be tuned by gate voltage. This is demonstrated in Figure \ref{fig:MR} (d) where we performed field rotation measurements at 2K at various gate voltages. Using similar analysis as in Figure \ref{fig:MR} (b)-(c), we find that the spin component increases in magnitude as electrons are depleted (negative gate voltage) from the 2DES (Figure \ref{fig:MR} (e)). As we elaborate in the discussion section, we relate this result to a change in position of the 2DES relative to the interface as function of carrier concentration.

While the results presented above point to magnetic fluctuations in the ETO layer scattering electrons in the 2DES, we find no evidence of the 2DES altering magnetism in the ETO layer, as one would expect from an RKKY-type magnetism. This can be seen in supplementary Figure S7, where susceptibility maps are measured at different back-gate voltages. We find that the susceptibility is independent on carrier density within resolution. 

To further examine the effect of conduction electrons on magnetism, we performed similar scanning SQUID measurements for non-ferroelectric LAO/ETO/STO hetereostructes. Two samples were studied - a conducting sample with a two-unit-cell thick layer of ETO at the interface and another sample with five-unit cells of ETO, which results in an insulating interface. Susceptibility versus back-gate voltage measurement were performed for the conducting sample, showing no effect of carrier modulation on the magnetic response (supplementary Figure S7). We note that our transport study reproduce previous results on the LAO/ETO/STO interface\cite{stornaiuolo2016tunable}  (supplementary Figure S5). The insulating sample studied here shows qualitatively similar results to the conducting samples (supplementary figure S6 (b)). These results further imply that the 2DES formed at these interfaces does not play a role in inducing magnetism in the thin ETO layer.

We now move to the question of the magnetic ground state of the thin ETO layer. For either a ferromagnetic or antiferromagnetic phase transition, one would expect a saturation of the magnetic susceptibility below the transition temperature. However, no such saturation is observed, suggesting that the ETO does not undergo magnetic ordering down to 1.7K. This observation is further supported by the zero-field spatial magnetic imaging produced at 1.7K, which shows no evidence of ferromagnetic domains (supplementary figure S2 (a)). An in-plane aligned magnetization is also ruled out, since fringing field lines from such domains would have been picked up at the presumed domain boundaries. These observations also hold for the non-ferroelectric LAO/ETO/STO interfaces studied here (see supplementary figures S2 (b) and S6 (a)).

No clear indication of a magnetic phase transition is visible in transport. This can be seen by the $ln(\frac{1}{T})$ upturn to the interface resistance, showing no saturation down $\sim$20mK. Typically, magnetic ordering is accompanied by a saturation of such a resistance upturn, as can be observed for example in systems such as doped ETO\cite{katsufuji1999transport,ahadi2017evidence,yamamoto2015antiferro,kususe2014magnetic} and manganites \cite{millis1998lattice,nagaev1996lanthanum}. Furthermore, no magnetic hysteresis is visible at 0.5K (supplementary figure S4), as opposed to, for example, the case of Sm doped ETO, which shows a strong hysteresis below the Curie temperature\cite{ahadi2017evidence}. 

\section*{Discussion}
In figure \ref{fig:Mainplot} (c) we sketch our understanding of how ferroelectricity and magnetism interact with the 2DES in the LAO/ETO/CSTO heterosctructure. The conduction electrons reside at the edge of the ferroelectric CSTO and are adjacent to the magnetic ETO layer. Ferroelectricity in the CSTO bulk can be switched by using the gate voltage to control the 2DES properties (Figure \ref{fig:RVG}). As the temperature is lowered, magnetic fluctuations in the ETO layer increase and scatter the conduction electrons in the 2D polar metal. This scattering results in a strong negative spin-MR signal (Figure \ref{fig:MR}) and a Kondo-type temperature dependence of the resistivity (Figure \ref{fig:Mainplot} (b), bottom panel) appearing together with a divering Curie-Weiss behaviour of the ETO susceptibility (Figure \ref{fig:SQUID} (c)). We note that similar Kondo-type temperature dependence of the resistance and negative MR were previously observed in doped ETO\cite{katsufuji1999transport,ahadi2017evidence,yamamoto2015antiferro,kususe2014magnetic} in manganites \cite{millis1998lattice,nagaev1996lanthanum} and in electron-doped cuprates \cite{finkelman2010resistivity}.

Furthermore, we show the spin MR term can be tuned by applying back-gate voltages, becoming stronger at negative gates (electron depletion, Figure \ref{fig:MR} (e)). We explain this enhancements by the 2DES becoming spatially closer to the ETO layer, making the 2DES more susceptible to magnetic scatterings. This picture of the electrons becoming more confined to the interface at the depleted regime is in line with the calculation of Delugas \etal\cite{delugas2011spontaneous}. 

While the 2DES experiences scattering off magnetic fluctuations originating in the ETO layer, our observations show that the magnetism in the ETO layer is unaffected by the presence of the 2DES. This conclusion is supported by: (1) The magnetic susceptibility is independent of gate voltage (supplementary figure S7). (2) Non-conducting ETO based heterostructures show qualitatively similar magnetic measurements to the conducting samples (supplementary figures S2 and S6). (3) All ETO-based samples studied here exhibit no ferromagnetism (supplementary figure S2 (a)-(c)).

Recently, LAO/STO interfaces were fabricated with GdTiO$_3$ interlayers instead of EuTiO$_3$\cite{lebedev2021gate}. These interfaces present similar, gate-tunable non-linear Hall signatures to those observed in the LAO/ETO/STO interface. However, no ferromagnetic domains were found in scanning SQUID measurements.

\section*{Summary}
In this work, we explored magnetism, ferroelectricity and their interactions with the 2D electron system formed at the the \LAO/\ETO/\CSTO~interface. The conducting electrons in the 2DES experience magnetic scattering originating from the \ETO~layer. This is evident from a diverging  $ln(\frac{1}{T})$ term in the interface resistance appearing at low temperatures along with a Curie-Weiss-type behaviour of the ETO susceptibility. By analysing magnetoresistance measurements, we were able to separate out the contribution of the spin scattering to the magnetoresistance. This scattering channel becomes dominant as temperature is lowered and is tunable by gate biasing, becoming stronger when electrons are depleted from the 2DES. Ferroelectricity also interacts with the 2DES as demonstrated by the hysteresis observed in sheet resistance when switching ferroelectric polarization in the bulk.

While the 2DES interacts with both ferroelectricity and magnetism, we find that the presence of conduction electrons has no effect on the magnetic susceptibility of the \ETO~layer. Furthermore, no signs of ferromagnetism is observed in any of our \ETO-based samples with or without ferroelectricity. Control of magnetism by electron density modulation is desired in order to realize a new type of multiferroic 2DES where ferroelectricity and magnetism can interact via an intimidating conducting layer. By selecting other magnetic materials it should be possible to achieve this desired functionality.

\section*{Methods}
\subsection*{Sample Preparation and Transport Measurements}
Two-unit cells of \ETO~followed by eight-unit cells of \LAO~were deposited by pulsed laser deposition on atomically flat TiO$_2$ terminated \CSTO~substrates. Both ETO and LAO were deposited in situ at a rate of 1 Hz, oxygen partial pressure of 10$^{-4}$ torr, temperature of 680C and energy density of 1.15 $\frac{J}{cm^2}$. After deposition, samples were cooled to room temperature at a rate of 3 $\frac{C}{min}$ at the deposition environment. The same growth procedure was followed for non-ferroelectric LAO/ETO/STO interfaces studied here, varying only the thickness of the ETO layer.
Back-gate electrodes were attached to the bottom of the CSTO with Ag paint. When gate voltage was applied, the leakage current was immeasurably small (<1 pA). The gate voltage is defined as positive when electrons accumulate at the interface. Measurements were performed in a PPMS system with a base temperature of 2K and magnetic fields up to 14T. Measurements below 2K were conducted in a Triton dilution refrigerator with a base temperature of 20 mK.

\subsection*{Scanning SQUID Measurements}
We use a scanning SQUID microscope with a micron size sensitive area (the pick-up loop). The pick-up loop is rastered above the surface of the sample, recording the Z component of the magnetic field as a function of position \cite{huber2008gradiometric,kirtley2012scanning,gardner2001scanning,persky2021studying}, in units of flux ($\Phi_{0}$). Local magnetometry, susceptometry, and current response measurements were performed simultaneously. Magnetometry mode maps the static magnetic landscape, captured by the SQUID’s pick-up loop. Susceptometry measurements are performed by applying local magnetic field with a field coil loop that surrounds the pick-up loop. In the data shown here a.c. current (0.1-3 mA RMS, 1-3 kHz) was applied to the field coil generates a local magnetic field of 0.1-3 G RMS. The pick-up loop detects magnetic signals generated by the applied field. This signal is separated from d.c. data by using a lockin. To compensate for background magnetic fields an identical pick-up loop, surrounded by an identical field coil, is located 1.2 mm away and wired in a gradiometric design. For mapping electric current flow, we capture the magnetic fields generated by the current by the pick-up loop. These signals are separated from others by using a lockin. A thread of current flow appears in a SQUID image as a negative magnetic signal next to a positive signal. We perform all scans at a fixed sensor-sample distance of $\sim 1 \mu m$.

\section*{Acknowledgements}
G.T. and S.W.S. contributed equally to this work. G.T. and Y.D. conceived the experiment, G.T and S.S fabricated samples, conducted magneto-transport and capacitance measurements. S.W.S and B.K. Performed the scanning-SQUID measurements. All authors analysed the data and wrote the manuscript. We thank Akira Yasuhara for conducting the STEM measurements. 
We thank Anshu Sirohi and Nadav Rotem for their help with the scanning SQUID measurements.

Work in TAU was supported by the Pazy Research Foundation grant No. 326-1/22, Israeli Science Foundation grant No. ISF-3079/20 and ISF-382/17 and The TAU Quantum Research Center. S.W.S. and B.K. were supported by the European Research Council Grant No. ERC-2019-COG-866236, the Israeli Science Foundation grant No. ISF-1281/17, COST Action CA16218, the QuantERAERA-NET Cofund in Quantum Technologies, Project No. 731473, and the Pazy Research Foundation grant No. 107-2018.
\clearpage

\onecolumn
\renewcommand\thefigure{S\arabic{figure}}    
\setcounter{figure}{0}  

\section*{Supplementary Material - Tunable Magnetic Scattering and Ferroelectric Switching at the \LAO/\ETO/\CSTO~Interface}
\begin{figure}[hbt!]
\centering
  \includegraphics[width=\columnwidth] {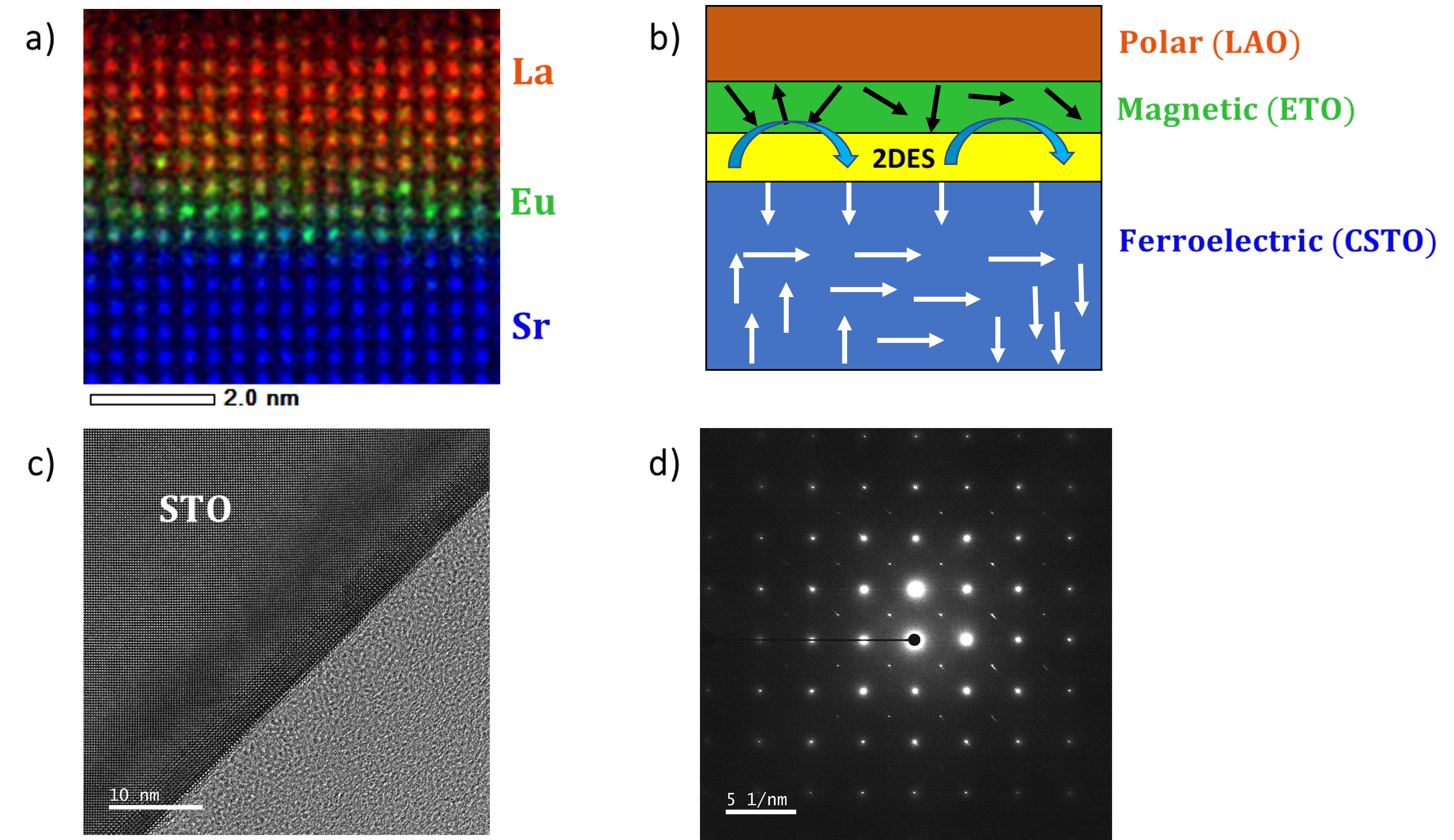}
  \caption{\textbf{Broad View of STEM Measurements} (a) STEM measurement along with EELS elemental identification confirming the high quality growth of the interface. (b) Sketch of our understanding of the system: The 2D polar metal is controllable by switching the bulk ferroelectricity and experiences scattering from spin fluctuations in the ETO layer. (a)+(b) also appear in the main text Figure 1. (c) Broad view of the heterostructure. (d) Diffraction data of the heterostructure showing high crystalline quality.} 
  \label{fig:S1}
\end{figure}
\clearpage

\begin{figure} [hbt!]
\centering
  \includegraphics{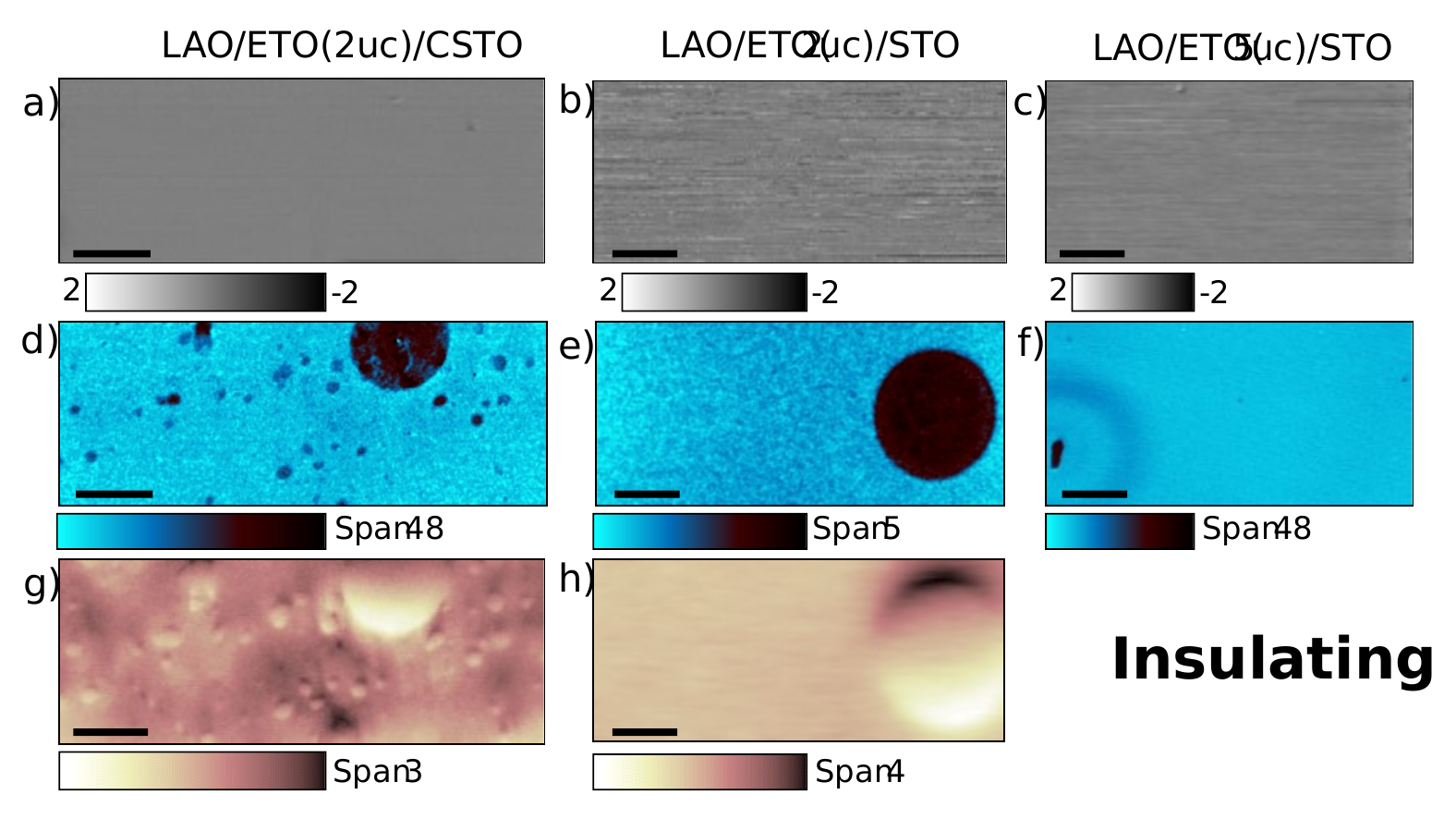}
  \caption{\textbf{Regions with reduced paramagnetic signal are also less conducting. The paramagnetic response is not accompanied by a ferromagnetic signal.} Scanning SQUID images of the three heterostructures studied in this work. LAO/ETO/CSTO (a,d,g), LAO/ETO(2 unit cell)/STO (b,e,h) and LAO/ ETO(5 unit cell)/STO (c,f), taken at 4.2K. (a-c) The static magnetic landscape (zero-field), no ferromagnetic signal is detected up to 2mG. (d-f) Magnetic susceptibility maps show paramagnetic response with isolated regions that are less (or not) paramagnetic. (g-h) Magnetic fields generated by the current flow enable tracking the spatial distribution of electric current. We find that regions that are less paramagnetic are also less conducting. Scale bars: 50µm;} 
  \label{fig:S2}
\end{figure}

\begin{figure} [hbt!]
\centering
  \includegraphics{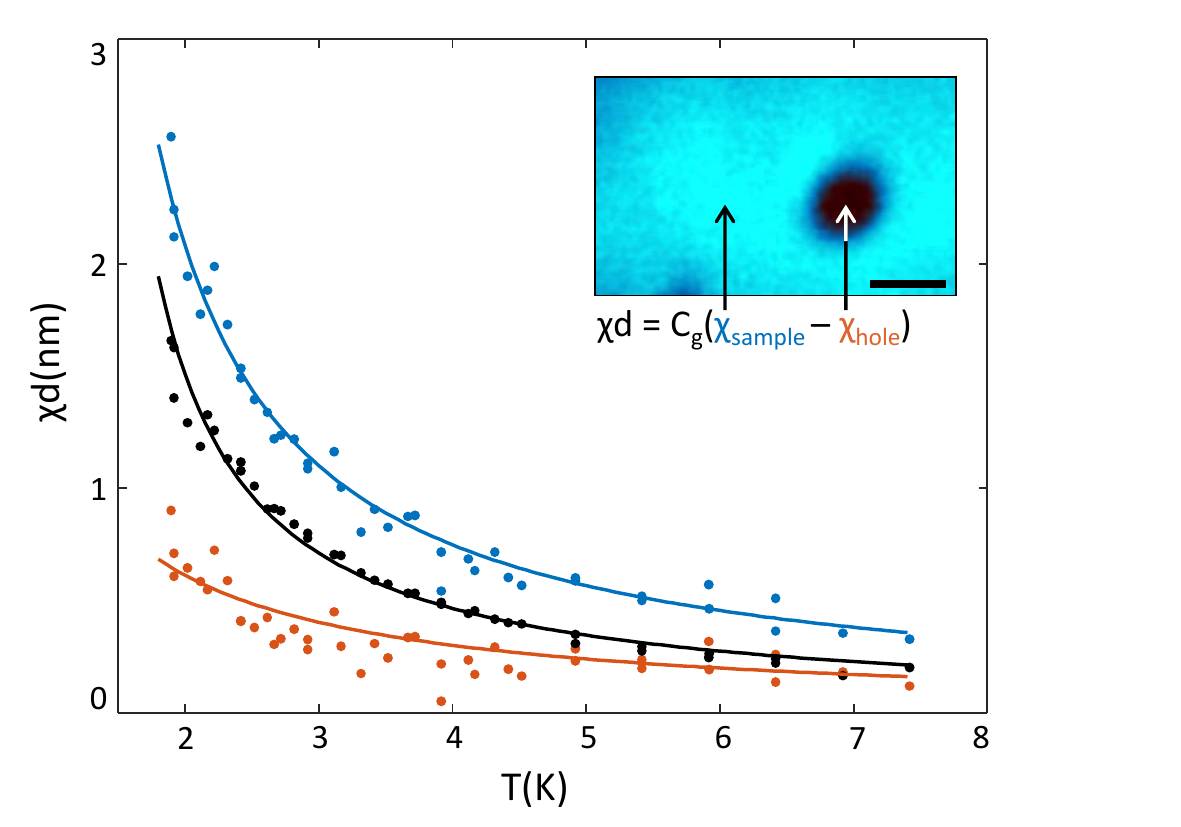}
  \caption{\textbf{Paramagnetic susceptibility decays with temperature.} Paramagnetic susceptibility as a function of the temperature is relatively uniform over $96.5\%$ of the sample (blue data). In order to track its temperature dependence we measure it relatively to a less paramagnetic region ("hole", dark circle in the inset, orange data). The difference between the value of the sample to the value of this specific hole represents a low limit of the decrease of the paramagnetic response with temperature (black data). Inset: Map of the paramagnetic response at 1.8K. Scale bar 10µm;} 
  \label{fig:S3}
\end{figure}
\begin{figure} [hbt!]
\centering
  \includegraphics[scale=0.5] {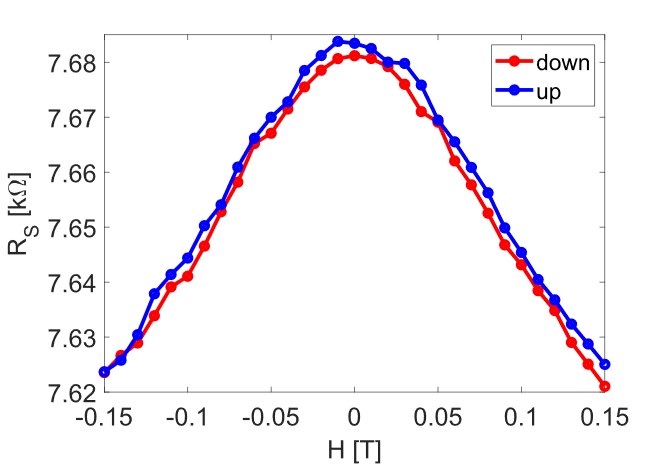}
  \caption{\textbf{No magnetic hysteresis in low temperature magnetoresistance} Interface resistance versus magnetic field of the LAO/ETO/CSTO heterostructure when sweeping magnetic field up and down at 0.5K. No hysteresis visible. Temperature was allowed to stabilize for 10min at each point before measuring as the 2DES resistance is extremely sensitive to temperature at this range.} 
  \label{fig:S4}
\end{figure}
\clearpage

\begin{figure} [hbt!]
\centering
  \includegraphics[width=\columnwidth] {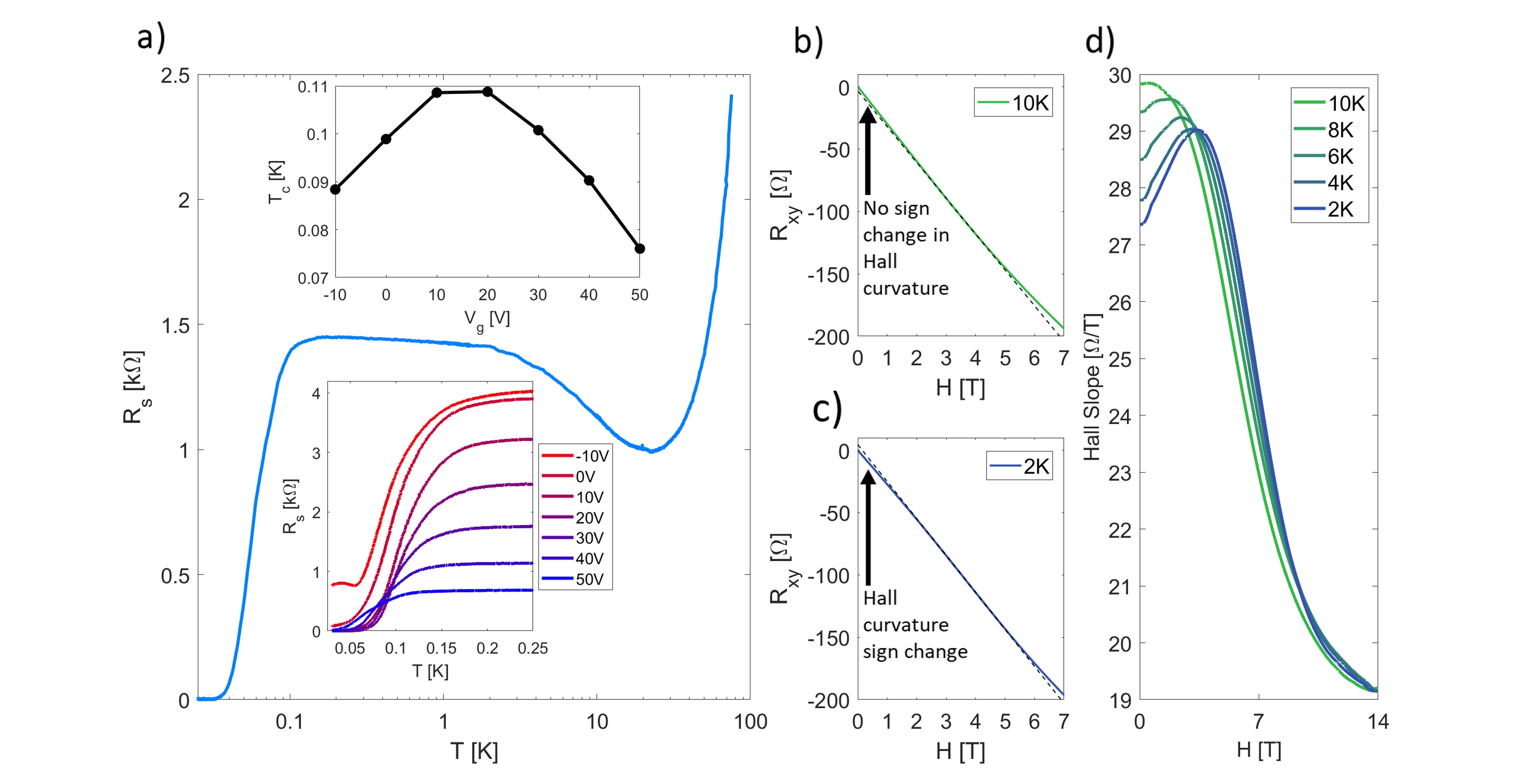}
  \caption{\textbf{Transport of the LAO/ETO/STO interface} (a) Resistance versus temperature in logarithmic scale. Upturn related to magnetic scattering seen at $\sim$20K and superconductivity at $\sim$100mK. Inset (bottom) superconductivity data at different back gate voltages. Inset (top) Superconducting $T_C$ versus gate showing dome behaviour. $T_C$ is defined as the temperature with half the normal state resistance. (b)-(c) Hall data at 10K and 2K respectively. At 10K the curvature of the hall does not change while at 2K it does. (d) Calculated derivative of Hall signal for different temperatures and fields. A clear curvature change is developed as temperature is lowered.} 
  \label{fig:S5}
\end{figure}
\begin{figure} [hbt!]
\centering
  \includegraphics{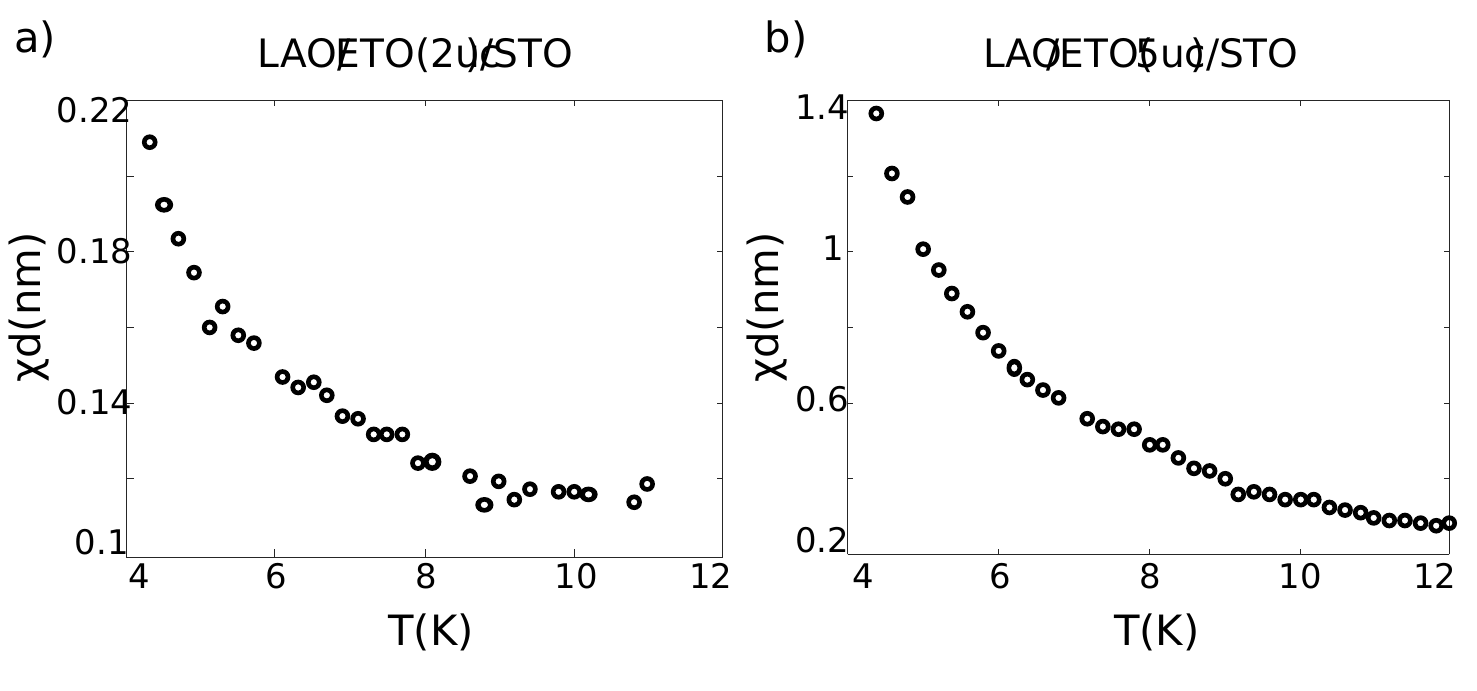}
  \caption{\textbf{The paramagnetic response scales with the thickness of the ETO layer.} Paramagnetic signals converted to susceptibility as a function of temperature on (a) LAO/ETO(2 unit cell)/STO heterostructure, and (b) LAO/ETO(5 unit cell)/STO.} 
  \label{fig:S6}
\end{figure}
\clearpage

\begin{figure} [hbt!]
\centering
  \includegraphics{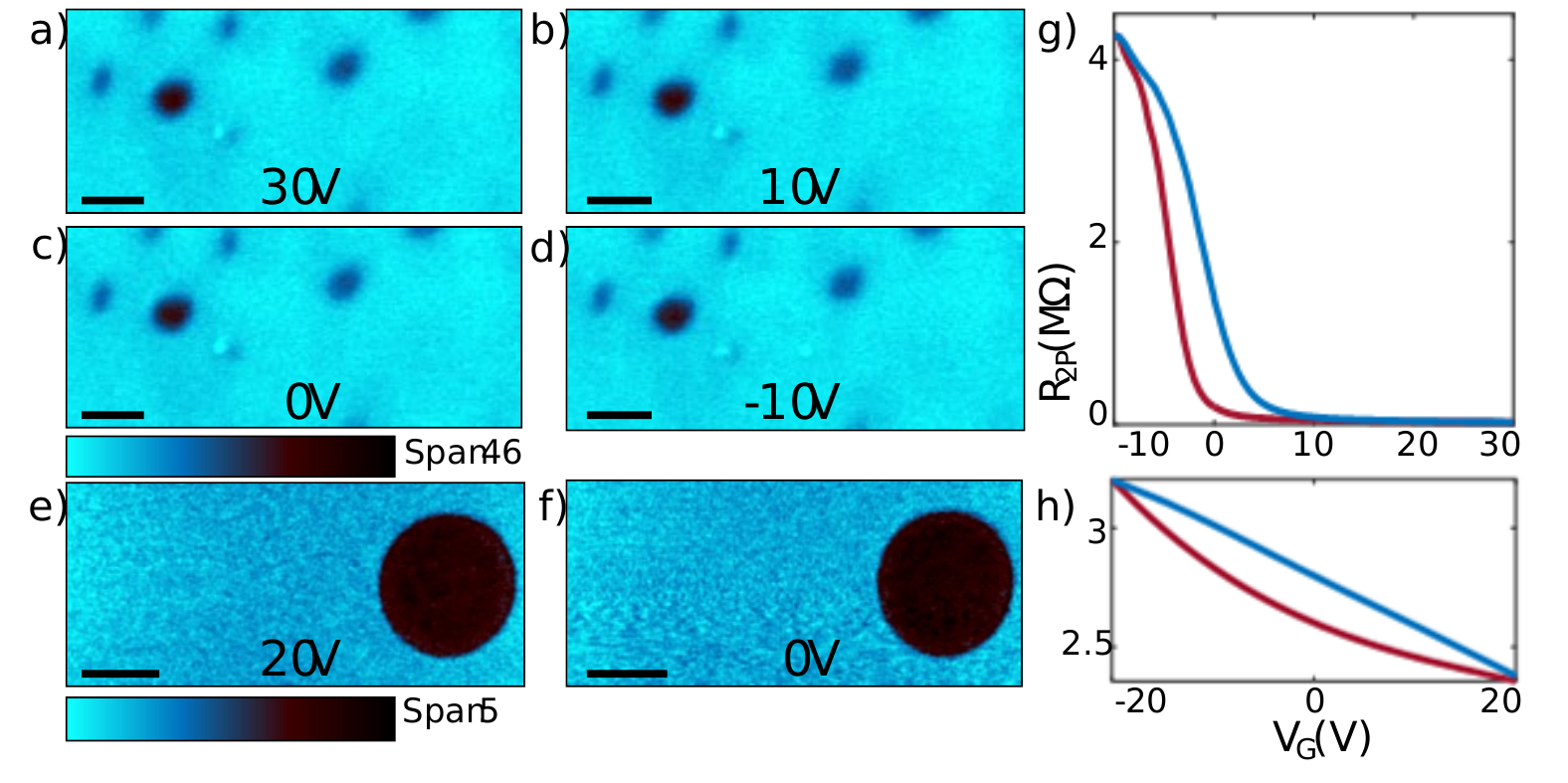}
  \caption{\textbf{The paramagnetic response does not depend on the back gate voltage.} (a-f) Maps of the paramagnetic response as a function of the back gate voltage. No observable difference between: (a) 30V (b) 10V (c) 0V (d) and -10V, at 1.8K in LAO/ETO(2 unit cell)/CSTO heterostructure, and (e) 20V and (f) 0V, at 4.2K in LAO/ETO(2 unit cell)/STO heterostructure. Scale bar b-e 20µm; g-h 50µm; (g-h) Two probe resistance as a function of back gate voltage shows hysteretic behavior at (g) LAO/ETO(2 unit cell)/CSTO heterostructure and (h) LAO/ETO(2 unit cell)/STO heterostructure.} 
  \label{fig:S7}
\end{figure}
\clearpage
\begin{figure} [hbt!]
\centering
  \includegraphics[width=\columnwidth] {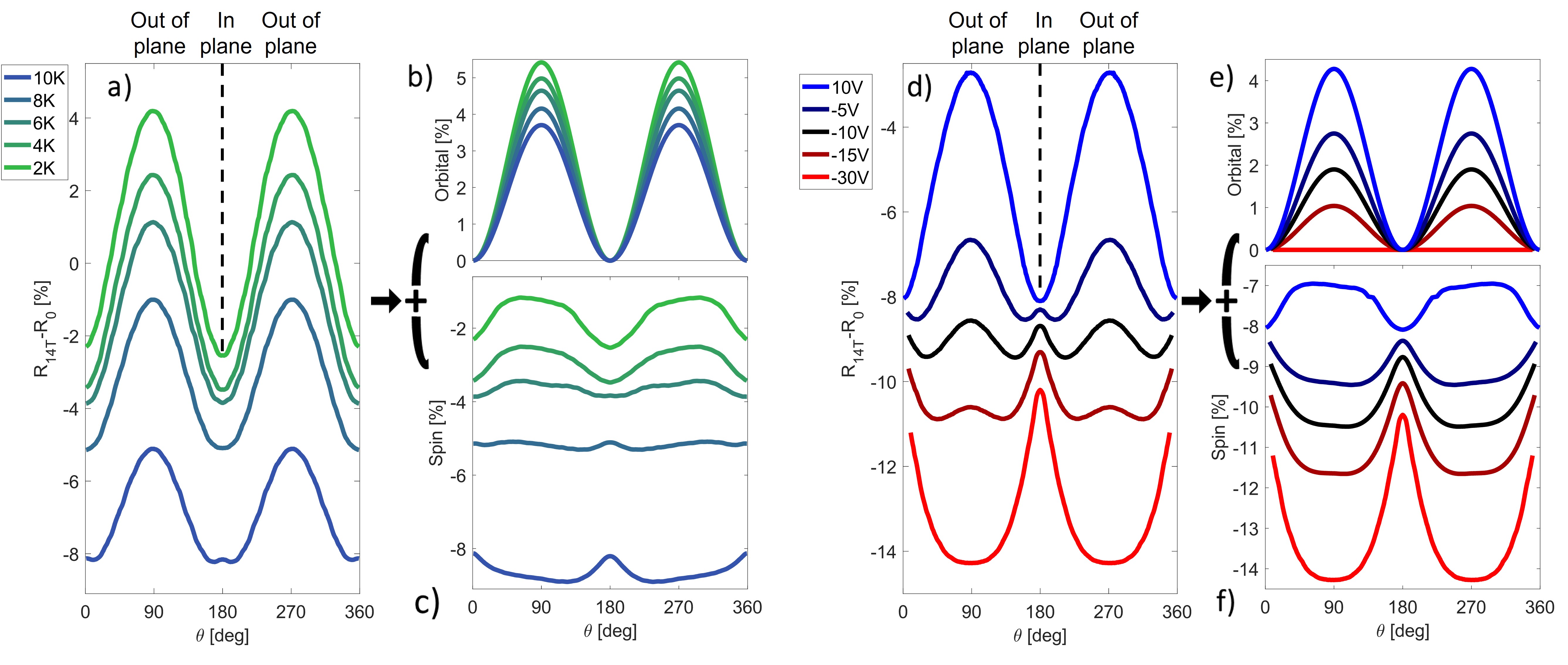}
  \caption{\textbf{Detailed Calculation of Orbital and Spin Components Versus Temperature and Gate Presented in Main Text Figure 4} (a) MR magnitude at 14T for different temperatures and angles between the interface and the field ($\theta$). A positive out of plane orbital contribution is observed at all temperatures. As temperature is lowered into the magnetic region, the entire MR signal becomes more negative. (b) Fitted orbital term for angles near the out-of-plane orientation ($70\degree\leq\theta\leq110\degree$)m (c) MR magnitude after subtraction of extrapolated orbital effect (b). This component represents the spin scattering term. (d) MR magnitude at 14T at 2K for different back-gate voltages. (e)-(f) Follow the procedure described in (b)-(c).} 
  \label{fig:S8}
\end{figure}
\clearpage

\section*{Appendix}
Converting SQUID measured magnetic flux by the pick-up loop ($\Phi_{PL}$) to unitless susceptibility ($\chi$)\cite{bluhm2009spinlike}:

$B=\mu_0\chi d\frac{dH}{dz}=\frac{\phi _{PL}}{\pi R_{PL}^2}$

Where $R_{PL}$ is the radius of the pick-up loop, and H is the magnetic field applied by the filed coil loop: 

$H=\frac{\mu_0R_{FC}^2}{2(z^2+R_{FC}^2)^{\frac{3}{2}}}I_{FC}$

$\frac{dH}{dz}=\frac{3\mu_0zR_{FC}^2}{2(z^2+R_{FC}^2)^{\frac{5}{2}}}I_{FC}$

$I_{FC}$ is the current applied in the field coil loop, $R_{FC}$ is the radius of the field coil loop, and z is the height of the SQUID pick-up loop above the sample. 

From these we can extract $C_g$ – the geometric constant for the conversion: 

$\chi d=\frac{2(z^2+R_{FC}^2)^{\frac{5}{2}}}{3\pi\mu _0zR_{FC}^2R_{PL}^2}\frac{\phi _{PL}}{I_{FC}}=C_g\frac{\phi _{PL}/\phi _0}{I_{FC}/A}$

For $z=1[\mu m] ; R_{FC}=6.3[\mu m] ; R_{PL}=1.25[\mu m]$ :

    $C_g=6*10^{-8}[m]$

$R_{PL}$ and $R_{FC}$ are the effective radius calculated from the inner and outer radiuses of the coils.

\end{document}